\documentclass[aip,amsmath,amssymb,reprint]{revtex4-1}
\usepackage{graphicx}
\usepackage{dcolumn}
\usepackage{bm}

\usepackage[utf8]{inputenc}
\usepackage[T1]{fontenc}
\usepackage{mathptmx}
\usepackage{etoolbox}
\usepackage{ulem}
\usepackage{siunitx}

\usepackage{xcolor}
\makeatletter
\def\@email#1#2{%
 \endgroup
 \patchcmd{\titleblock@produce}
  {\frontmatter@RRAPformat}
  {\frontmatter@RRAPformat{\produce@RRAP{*#1\href{mailto:#2}{#2}}}\frontmatter@RRAPformat}
  {}{}
}%
\makeatother
\begin{document}

\preprint{AIP/123-QED}

\title{Conformal-invariance of 2D quantum turbulence in an exciton-polariton fluid of light}

\author{R. Panico}
\affiliation{CNR NANOTEC, Institute of Nanotechnology, Via Monteroni, 73100 Lecce, Italy}

\author{A. S. Lanotte}
\affiliation{CNR NANOTEC, Institute of Nanotechnology, Via Monteroni, 73100 Lecce, Italy}
\affiliation{INFN, Sez. Lecce, 73100 Lecce, Italy}

\author{D. Trypogeorgos}
\affiliation{CNR NANOTEC, Institute of Nanotechnology, Via Monteroni, 73100 Lecce, Italy}

\author{G. Gigli}
\affiliation{CNR NANOTEC, Institute of Nanotechnology, Via Monteroni, 73100 Lecce, Italy}
\affiliation{Dipartimento di Matematica e Fisica E.~De Giorgi, Universit\`a del Salento, Campus Ecotekne, via Monteroni, Lecce 73100, Italy}

\author{M. De Giorgi}
\affiliation{CNR NANOTEC, Institute of Nanotechnology, Via Monteroni, 73100 Lecce, Italy}

\author{D. Sanvitto}
\affiliation{CNR NANOTEC, Institute of Nanotechnology, Via Monteroni, 73100 Lecce, Italy}

\author{D. Ballarini}
\email{dario.ballarini@nanotec.cnr.it}
\affiliation{CNR NANOTEC, Institute of Nanotechnology, Via Monteroni, 73100 Lecce, Italy}

\date{\today}

\begin{abstract}
The similarities of quantum turbulence with classical hydrodynamics allow quantum fluids to provide essential models of their classical analogue, paving the way for fundamental advances in physics and technology. Recently, experiments on 2D quantum turbulence observed the clustering of same-sign vortices in strong analogy with the inverse energy cascade of classical fluids. However, self-similarity of the turbulent flow, a fundamental concept in the study of classical turbulence, has so far remained largely unexplored in quantum systems. Here, thanks to the unique features of exciton-polaritons, we measure the scale invariance of velocity circulations and show that the cascade process follows the universal scaling of critical phenomena in 2D. We demonstrate this behaviour from the statistical analysis of the experimentally measured incompressible velocity field and the microscopic imaging of the quantum fluid. These results can find wide application in both quantum and classical 2D turbulence.   
\end{abstract}

\maketitle

\section{\label{sec:1}INTRODUCTION}

Turbulent dynamics in classical fluids has been first identified by A. N. Kolmogorov with the presence of a self-similar energy cascade. Self-similarity, or scale invariance, refers to the statistically identical behaviour of the velocity fluctuations after scale transformations. In three dimensions (3D), the existence of non-zero energy dissipation even in the limit of zero viscosity implies a direct energy cascade from a large injection scale towards smaller spatial scales~\cite{Frisch1995}. In 2D, the picture is different due to the existence of an additional integral of motion given by enstrophy.
%, defined as the squared vorticity $\left|\mathbf{\omega}\right|^2$, with $\mathbf{\omega}=\nabla\times\mathbf{u}$ the vorticity field and $\mathbf{u}$ the incompressible velocity. 
Since energy and enstrophy can not cascade in the same direction~\cite{Fjo1953}, in 2D the kinetic energy flows from the injection scale towards larger structures (inverse energy cascade), while enstrophy undergoes a direct cascade toward smaller scales~\cite{kraichnan1967, boffetta2012two}. 

A subtler consequence of the direction of the cascade concerns the presence of intermittency, that is the deviation from self-similarity. Indeed, direct turbulent cascades are generally not scale invariant~\cite{AlexakisBiferale2018}, in connection with the non-gaussian nature of the small-scales viscous processes. This is the case for 3D turbulence, where most of the kinetic energy is dissipated within spatially localised structures. This intermittent statistics of energy dissipation breaks the original Kolmogorov assumption of self-similarity of the cascade process~\cite{Yeung2015,benzi1984multifractal}. 
The 2D energy cascade in classical turbulence is fundamentally different: the presence of intermittency in the inverse transfer process has been ruled out\cite{ParetTab1998,boffetta2000inverse,boffetta2012two}, intuitively because the energy fluxes are directed towards larger structures where extreme events due to energy dissipation play no role. Hence velocity fluctuations are scale-invariant in the inverse energy cascade. %Moreover, in 2D turbulence, scale invariance can be promoted to conformal invariance, and it has been suggested that the inverse cascade manifests the same universal behaviour of critical percolation~\cite{Bernard2006}%Moreover, it has been shown that the isovorticity lines in the inverse energy cascade shares the same statistical behaviour of critical percolation, suggesting a deeper connection between the scale invariance of turbulent flow and fundamental models of phase transitions~\cite{Bernard2006}.

Quantum turbulence, differently from its classical counterpart, is intrinsically singular, since its basic constituents are discrete, quantised vortices of unitary topological charge. Quantum vortices resemble the point vortex model proposed by L. Onsager, who described the final stage of the inverse cascade as an equilibrium state in a negative temperature regime~\cite{onsager1949statistical,Eyink2006}. The experimental investigation of quantum turbulence, which began with superfluid helium~\cite{HallVinen56,Barenghi2014}, has made great progress with the realisation of Bose-Einstein condensates (BEC) of ultracold atoms~\cite{hennPRL2009,white2014, navon2016, navon2021quantum}. One of the main reasons why BECs are important for the study of quantum turbulence is the possibility of visualising individual vortices and describing their dynamics on a microscopic level~\cite{kwon2021sound}.
%Quantum vortex dynamics is described at the %microscopic level by the Gross-Pitaevski equation, %which, after Madelung transformation, allows a direct %hydrodynamic interpretation as a slightly modified %Euler equation. The difference is the presence of %quantum pressure, which is responsible for vortex %reconnections with emission of sound pulses and for %the nucleation of vortices near a boundary or a %strong density variation. 
On the other hand, an open question is how to bridge the gap between the discrete picture of quantum vortices and the self-similar nature of classical inverse energy cascade~\cite{johnstone2019evolution,gauthier2019giant,Skaugen2017}. In particular, we wonder whether the self-similar spatial correlations may nonetheless emerge in the velocity field of a quantum fluid, and which are the roots of this evidence in the system dynamics.

For this purpose, we use an optical system, exciton-polaritons in semiconductor microcavities. These hybrid light-matter quasiparticles have been shown in the last decades to behave like a quantum fluid of light, manifesting out-of-equilibrium Bose-Einstein condensation and superfluidity~\cite{carusotto2013quantum, amo2009collective, caputo2018topological, ballarini2020directional}. We have recently shown that, under suitable initial conditions, it is possible to induce an inverse cascade of incompressible kinetic energy in 2D polariton quantum fluids~\cite{panico2023onset}.
%The main advantage of these systems is the high accuracy in measuring the velocity field of a large number of quantum vortices, which allows for a robust statistical analysis~\cite{donati2016twist,panico2021dynamics}. 
Crucially, these systems allow the measurement of the spatial distribution of the velocity field with high accuracy, enabling a robust statistical analysis~\cite{donati2016twist,panico2021dynamics}. 
 %Quantised vortices, injected in the system as dipoles at the scale of the vortex core, organise in clusters at larger scales where the quantum nature of the  polariton fluid becomes less important, and a regime comparable to classical Kraichnan-Kolmogorov inverse cascade emerges. The signature is in the scaling behaviour of the incompressible kinetic energy power spectrum behaving as $E(k) \propto k^{-5/3}$, in a suitable range of the Fourier spectrum.
 
Here, thanks to the direct measurement of the phase of the polariton field, we are able to extract the statistics of the incompressible velocity in the inverse energy cascade of a quantum fluid. While the singular nature of quantised vortices manifests itself in the large tails of velocity increment distribution, the velocity circulations (or vorticity fluxes) show remarkable scale invariant properties. Moreover, we find that a coarse graining of the vorticity field allows the identification of macro-regions of aligned vortex-antivortex dipoles that are responsible for the appearance of long-range order in the system. We show that the statistical distribution and fractal dimension of the regions with correlated vorticity follow the critical behaviour predicted near phase transition in percolation theory. This analysis goes beyond the classification of first-neighbour vortices and opens to the investigation of spatial symmetries in 2D quantum turbulence.

\begin{figure}
    \centering
    \includegraphics[width=\columnwidth]{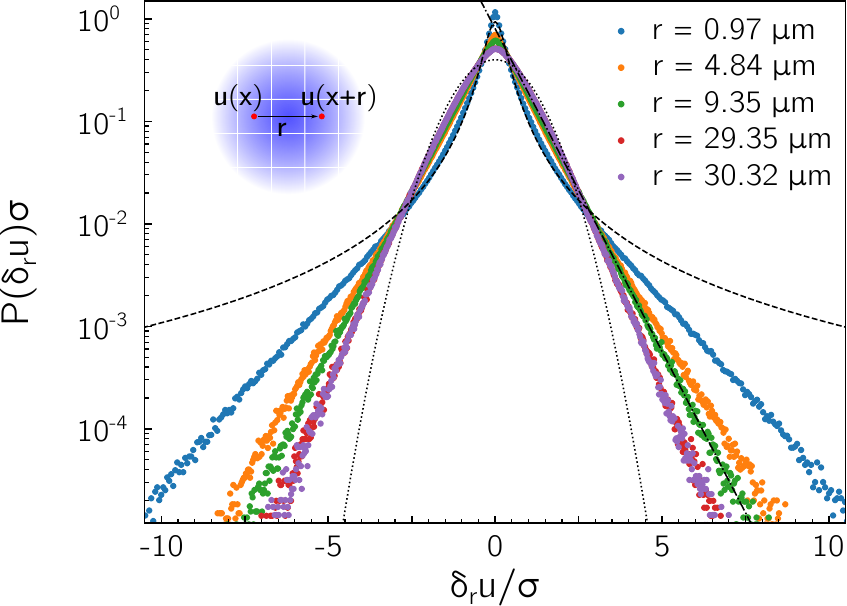}
    \caption{a) Probability density functions (PDF) of the longitudinal velocity increments $\delta_r u \equiv\frac{\mathbf{r}}{r}\cdot\left[\mathbf{u}(\mathbf{x}+\mathbf{r})-\mathbf{u}(\mathbf{x})\right]$ at scale $\mathbf{r}$, normalised to their standard deviation $\sigma=\langle (\delta_r u)^2 \rangle^{1/2}$. For sub-healing length scales, $r < \xi=2.3$~\si{\micro\meter}, the PDF follows a Cauchy-Lorentz distribution (dashed line), while at increasing $\mathbf{r}$ it slowly approaches a Gaussian (dotted line). For every scale, the tails of the distribution are exponential (dashed-dotted line). The lack of a self-similar rescaling of the PDFs is an indication of intermittency.
    %b) Log-log plot of the second order moment of the incompressible velocity increments as a function of the scale. A scaling behaviour $S_2(r) \sim r^{\zeta_2}$ is expected within the turbulent scaling range: a slope is fitted in the scaling region of few healing lengths, which results in the value of the scaling exponent $\zeta_2=0.7\pm0.02$. c) Values of the exponents $\zeta_p$, obtained as in (b) for various moments $S_p(r)$, in the same spatial interval, show a deviation from self-similarity (dashed curve).
    }
    \label{fig:1}
\end{figure}

\section{\label{sec:2}MAIN RESULTS}   
We use the same experimental configuration described in a previous work~\cite{panico2023onset} to induce vortex clustering and a transient regime of inverse kinetic energy cascade after the expansion of a polariton quantum fluid in a confining potential. The typical number of vortices is N>80, with intervortex distances $\approx2\xi$ and diameter of the confining ring potential $>50\xi$. For our analysis, we consider the incompressible velocity field $\mathbf{u}(\mathbf{x},t)$ of the 2D polariton fluid described by the wave-function $\psi(\mathbf{x},t)=\sqrt{\rho(\mathbf{x},t)}e^{-i\theta(\mathbf{x},t)}$, in the time interval corresponding to the inverse energy cascade. The total velocity field is given by the gradient of the phase $\theta$ as 
\begin{equation}
    \mathbf{v}(\mathbf{x},t)=\frac{\hbar}{m}\nabla\theta(\mathbf{x},t),
\end{equation}
where $m$ is the polariton mass. The phase of the polariton fluid is obtained from time-resolved interferometric measurements of the photons emitted by the microcavity, while the incompressible component $\mathbf{u}(\mathbf{x},t)$ of the velocity field is determined at any point in space by applying the Helmholtz decomposition to the density weighted velocity $\sqrt{\rho}\mathbf{v}$\,~\cite{panico2023onset}. 
We use about $125 K$ pixels per image with a spatial resolution $s\sim0.14\xi$, being $\xi=2.3$~\si{\micro\meter} the estimated healing length of the quantum fluid. To improve statistical accuracy, we average over larger time intervals (always within the inverse energy cascade window), and over few realisations of the experiment in different locations of the sample. In the following, we show the statistics over a total number of pixels $\approx 2\cdot10^7$ (accounting for the longitudinal increments along both x and y axis). 

%\ric{\sout{The probability density function (pdf) of the velocity increments $ S_{p}\left(r\right)$ with $p=1$, normalised by their standard deviations, are calculated as}
% \begin{equation}
%     A=B.
% \end{equation}}
%
In Fig.~\ref{fig:1}, the probability density function (PDF) of the longitudinal velocity increments $\delta_r u \equiv\frac{\mathbf{r}}{r}\cdot\left[\mathbf{u}(\mathbf{x}+\mathbf{r})-\mathbf{u}(\mathbf{x})\right]$, where the velocity ${\bf u}$ and the separation ${\bf r}$ vectors are taken in the same direction, are shown for increasing spatial scale $r=|{\bf r}|$, from $r\simeq0.4\xi$ to $r\simeq 14\xi$. %As shown in the SI, the pdf have a positive skewness at scale above the healing length, confirming the directional flux of energy toward larger scales as expected for the inverse energy cascade~\cite{bofetta2000inverse}.
At very small scales, the velocity increment PDF follows a Cauchy distribution, while it gradually approaches a Gaussian distribution as the scale increases~\cite{min1996levy}. The presence of exponential tails in the distribution is clearly visible for every distance, indicating a finite probability of large $\left|\delta_r u\right|$ events. This follows directly from the presence of quantised vortices, where velocity increments manifest their singular behaviour~\cite{min1996levy}. This reflects as well in the lack of self-similarity of the PDFs when rescaled to their standard deviations as in Fig.~\ref{fig:1} -- self-similarity would entail a collapse of the PDFs onto a single curve, as observed in the classical case~\cite{boffetta2000inverse}. This intermittent behaviour is intrinsic to the quantum system and can be related to the fact that energy is injected spontaneously through the nucleation of vortex dipoles at the healing length scale~\cite{sofiadis2023inducing}.% Incidentally, we remark that at difference with the locality regime of classical inverse cascade, the scaling behaviours of velocity and vorticity cannot be directly linked in a quantum system.

\begin{figure}
    \centering
    \includegraphics[width=\columnwidth]{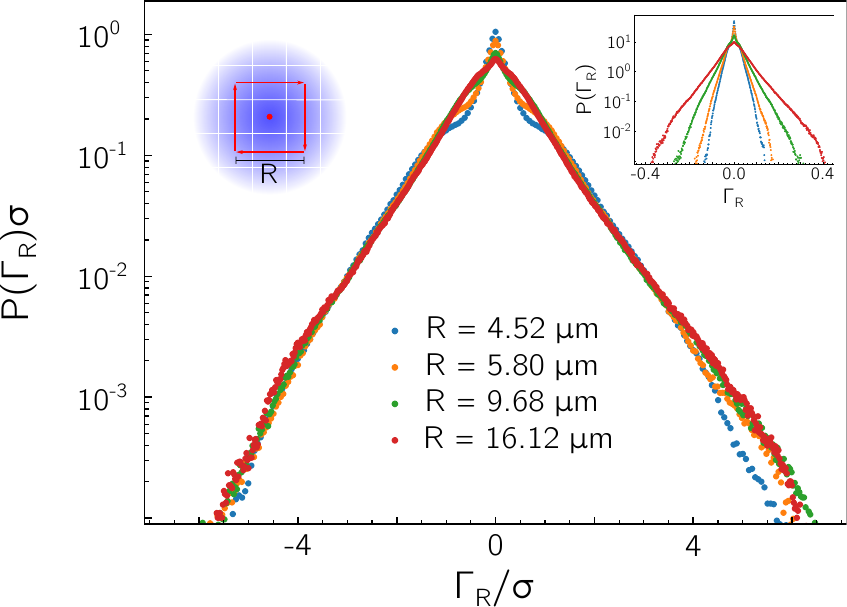}
    \caption{PDF of the velocity circulation for loops of size $R$, defined as Eq.\ref{circulation}, normalised to its standard deviation $\sigma\equiv \langle \Gamma_R^2 \rangle^{1/2}$. The inset shows the PDFs for the velocity circulations $\Gamma_R$ without rescaling.}
    \label{fig:2}
\end{figure}

While the distribution of the velocity increments is not a good observable to look for scale-invariant properties in a quantum fluid, moments of velocity circulation has recently been proposed as a more fundamental and unifying quantity~\cite{Iyer2019,Polanco2021,zhu2023circulation}. Velocity circulation is defined as
\begin{equation}
    \Gamma_R(\mathbf{x})=\oint_{C_R}\mathbf{u}(\mathbf{x}')\cdot\mathrm{d}\mathbf{x'},
\end{equation}
where $C_R$ is a square loop with opposite corners $(x-R/2,y-R/2)$ and $(x+R/2,y+R/2)$, while $\mathbf{u}$ is the incompressible velocity. 
In Fig.~\ref{fig:2}, we show the PDFs of the velocity circulation, measured on closed loops of size $R$. The most striking feature is that scale invariance now holds: contrary to the case of velocity increments, the PDF of the velocity circulation on loops of different size rescale with their standard deviation and collapse on a single (exponential) distribution. Incidentally, we remark that at difference with the locality regime of classical inverse cascade, the scaling behaviours of velocity and vorticity cannot be directly linked in a quantum system.
%We will further exploit this property to propose a conformal %invariant evidence of the vorticity in the system.
%
\begin{figure}
    \centering
    \includegraphics[width=\columnwidth]{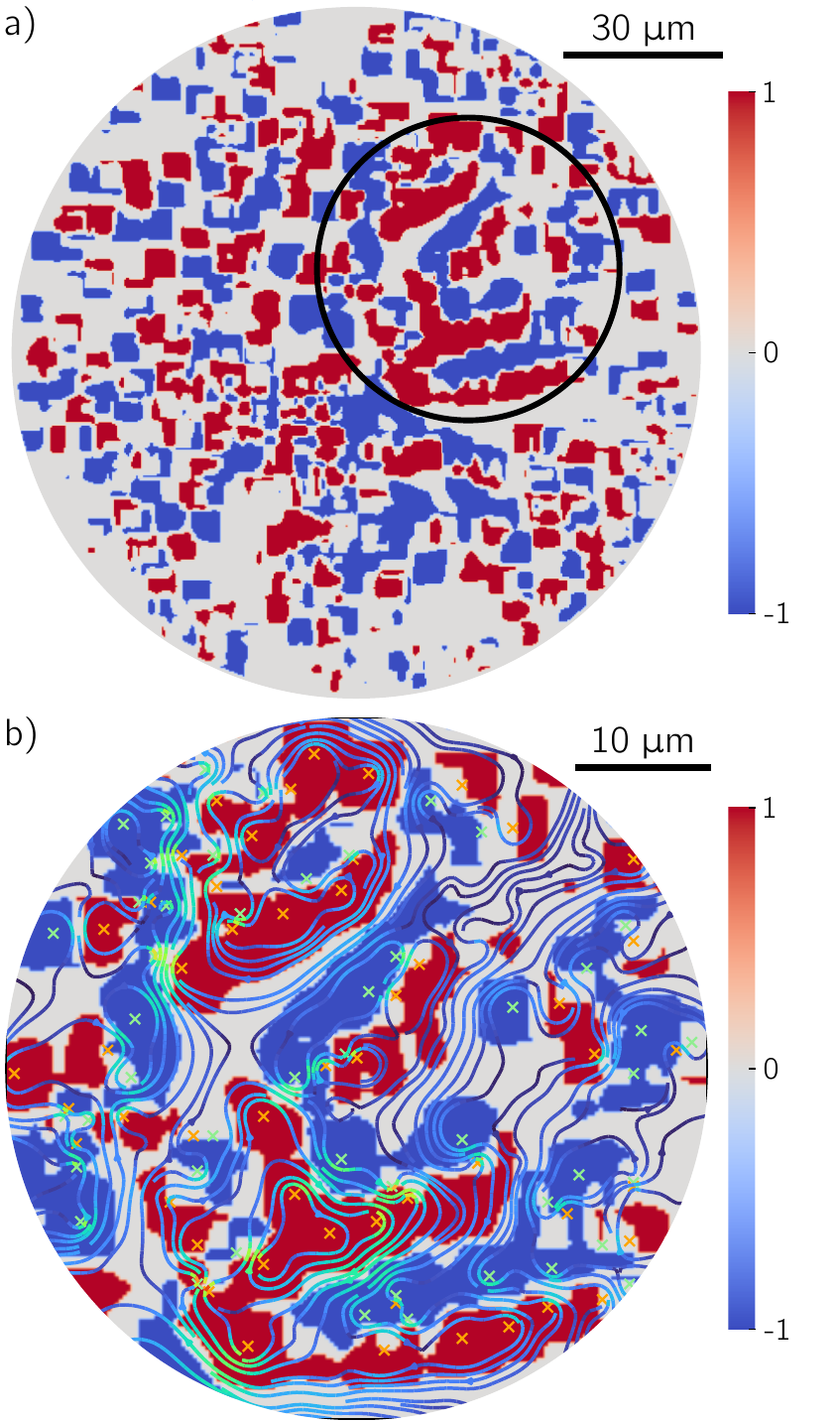}
    \caption{a) Circulation map of a single frame obtained using Eq.~\ref{circulation} with $R\simeq\xi$, showing the connected areas with a circulation value above 10\% of its maximum. b) Zoom of the spatial region (black circle in a), to better contrast areas of iso-vorticity flux with respect to the velocity field streamlines; the positions of the vortices (orange) and antivortices (green) are also reported. The overlap between closed velocity lines and the areas we identified is considerable and directly related to specific orientation of the dipoles.}
    \label{fig:3}
\end{figure}

\begin{figure}
    \centering
    \includegraphics[width=\columnwidth]{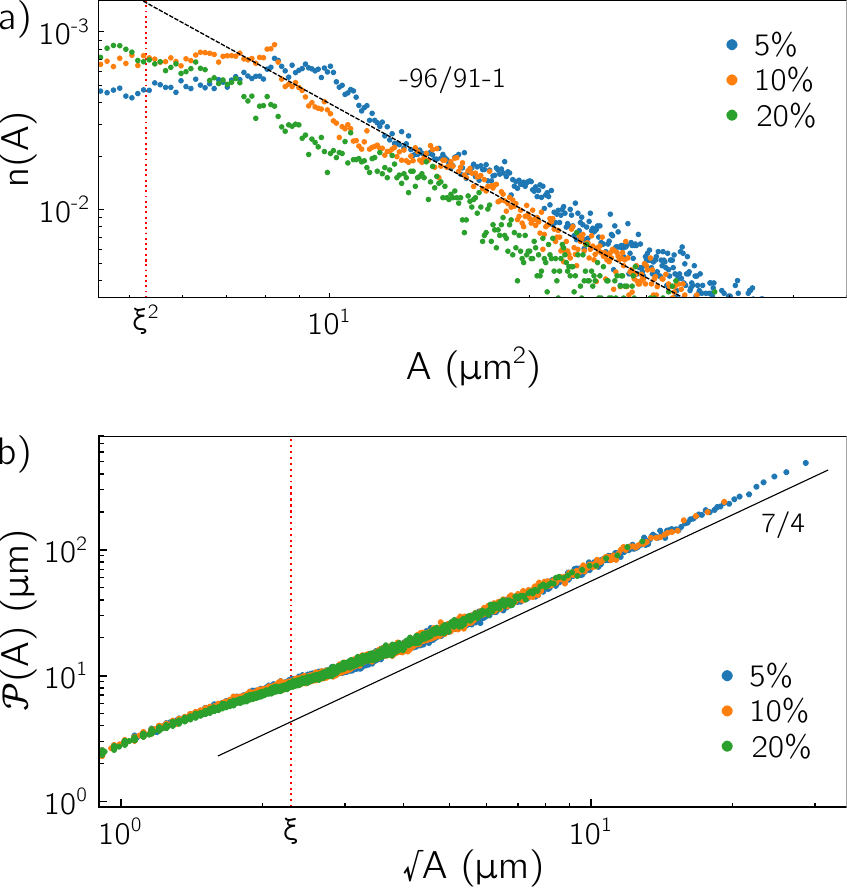}
    \caption{a) Distribution of the extension of the connected areas of iso-vorticity flux; threshold values ($5\%, 10\%,20\%$) are referred to the maximum value. The distribution scales with the exponent $\alpha\simeq96/91 + 1\simeq2.05$ for values of the area above $\xi^2$. b) Fractal dimension of the connected regions computed as the perimeter of each region versus their size, i.e. the square root of their area, yielding a value of $D=7/4$.}
    \label{fig:4}
\end{figure}
Self-similarity is a strong feature that suggests the quest for other global symmetries. Numerical results on classical 2D turbulence have shown that lines of zero-vorticity, i.e. the boundaries of high vorticity regions, are stochastic Schramm-Loewner evolution ($SLE_\kappa$) curves~\cite{Bernard2006}. The SLE curves univocally describe the conformal invariant scaling limit of the interfaces of many 2D critical models, with the parameter $\kappa$ defining the universal behavior close to threshold. The same universality class was predicted for cluster boundaries of critical percolation, one of the most fundamental models of phase transition, and the isovorticity lines in the inverse energy cascade, also for weakly compressible fluids~\cite{puggioni2020conformal}.

To explore the appearance of conformal invariant interfaces in 2D quantum turbulence, we consider a continuous field of coarse-grained vorticity $\mathbf{\omega}=\nabla\times\mathbf{u}$ (i.e. its flux), defined at each point in space as the value of velocity circulation $\Gamma_R(\mathbf{x})$ over a square loop of size comparable to the healing length $\xi$. %We therefore calculate for each point $\mathbf{x}=(x,y)$
%\begin{equation}
 %   \Gamma_R(\mathbf{x})=\oint_{C_R}\mathbf{u}(\mathbf{x}')\cdot\mathrm{d}\mathbf{x'},
%\end{equation}
%where $C_R$ is a square loop with opposite corners $(x-R/2,y-R/2)$ and $(x+R/2,y+R/2)$.
Via the Stokes' theorem, we calculate for each point $\mathbf{x}=(x,y)$ the vorticity flux through the surface $S_R$ enclosed in the loop $C_R$ as
\begin{equation}
\Gamma_R(\mathbf{x})=\int_{S_R}\left(\nabla\times\mathbf{u}(\mathbf{x'})\right)\cdot\mathrm{d}\mathbf{S}_R(\mathbf{x'}).
\label{circulation}
\end{equation}
The scalar field $\Gamma_R(\mathbf{x})$ combines the information about the sign and distribution of dipoles, smoothing the singularities of quantum vortices. The 2D maps of $\Gamma_R(\mathbf{x})$ are binarized using an upper (lower) threshold to identify the areas with correlated vorticity (see methods). In Fig.~\ref{fig:3}a, we show a typical 2D map for $\Gamma_{2\xi}(\mathbf{x})$ with a threshold of 10\% of its maximum value.
%%%
In Fig.~\ref{fig:3}b, the region within the black circle in Fig.~\ref{fig:3}a is magnified to show the underlying organization of vortices and velocity streamlines. The regions with same sign of $\Gamma_{2\xi}(\mathbf{x})$ are indicated in red and blue, for positive and negative fluxes, respectively. Vortices and antivortices are also indicated in Fig.~\ref{fig:3}b, showing the close relation between dipole orientation and the formation of extended regions with similar vorticity. While a first neighbour classification would only count the presence of a large number of dipoles, here we identify clusters of aligned vortex dipoles forming coherent structures that extend over several healing lengths.%This method appears then well suited for the study of the formation of large scale order. %and goes beyond the typical vortex classification which is limited to a nearest neighbours approach.
% gia' detto varie volte
%
%In 2D classical turbulence, it has been noted that the isovorticity lines can be mapped to the random clusters obtained in percolation models close to the critical threshold~\cite{Bernard2006}. The stochastic Schramm–Loewner evolution (SLE) technique, which univocally maps the scaling limit of critical models to conformal invariant transformations, allows to estimate the same universality class for 2D inverse energy cascade and critical percolation. In particular, SLE curves are parameterised by the diffusivity parameter $\kappa$ of Brownian motion, with $\kappa=6$ for percolation. One of the predictions for SLE curves is their fractal dimension, which is known to be $D = 1+\kappa/8$. 
%

We focus on the statistics of the vorticity (flux) clusters and on that of their boundaries, for which percolation theory predictions have already been tested in classical turbulence~\cite{Bernard2006,puggioni2020conformal}. 
%It is interesting therefore to compare the statistics of the clusters of aligned dipoles measured in our 2D quantum turbulence setup with the conformal invariant predictions of classical turbulence. 
%
In Fig.~\ref{fig:4}a, we show the distribution $n(A)$ of regions with area $A$, identified as in Fig.~\ref{fig:3}b. The PDF decreases as a power law $n(A)\propto A^{-\alpha}$, with the same exponent $\alpha$ independently from the threshold used. The dashed black line scales as $\propto A^{-96/91 -1}$ and characterises the cluster size distribution at the percolation threshold~\cite{book_percolation}, showing good agreement with the experimental results above the healing length. In the absence of the energy cascade, we do not observe such clear scaling of the size distribution, as shown in the SI. 
%For lengths above $\approx1.3\xi$, it shows a good %agreement with our experimental results.

One of the central predictions for SLE curves is their fractal dimension, which is known to be $D = 1+\kappa/8$ and, for percolation, $\kappa=6$ and $D=7/4$. In Fig.~\ref{fig:4}b, we plot the perimeter $\mathcal{P}(A)$ versus the square root of the intrinsic area $A$ of the connected regions. A clear scaling $\mathcal{P}(A)\propto (\sqrt{A})^{D}$ appears, with the exponent $D$, indicating the fractal dimension, which is $D>1$. The black line in Fig.~\ref{fig:4}b is the SLE prediction for percolation, $D=7/4$, which shows a good agreement with the experimental results above the healing length. In the absence of the inverse energy cascade, the perimeter-area scaling is limited by the smaller spatial extension of the isovorticity regions (see SI), which have maximum size of $\approx2.5\xi$ as compared to more than $10\xi$ in Fig.~\ref{fig:4}b.

\section{\label{sec:3}CONCLUSIONS}
To our knowledge, this is the first experimental measurement of the statistics of the velocity field in turbulent 2D quantum fluids. %The landmark of turbulence is the Kolmogorov spectrum and this implied that historically the scaling properties of the kinetic energy have been mostly analysed in the Fourier plane. 
%With the ability to follow the dynamics of quantum vortices in atomic BECs, theoretical models in terms of a microscopic description of quantised vortices have been tested with respect to real space analysis. Nevertheless, the pioneering observation of clustering of equal sign vortices in 2D BECs has involved a limited number of vortices only. More generally, the observation of the global symmetries of quantum turbulence is still an endeavouring task.
%More recently, the growing interest in quantum fluids of light lead to the development of new techniques to track the vortex dynamics and the measure the velocity field with high spatial resolution, providing an alternative platform which can be insightful in this direction. 
%
%Moreover, the turbulent dynamics is triggered after the expansion of the polariton fluid against the potential barrier, cancelling out any potential trace of the pulsed excitation from the subsequent dynamics~\cite{panico2023onset}. %We also note that, for exciton-polaritons, the field is exactly 2D, the confinement of the cavity in the third direction being over a much smaller distance than the healing length.
Our experiments show that the scale invariance of classical turbulence can be retrieved by considering velocity circulation instead of velocity increments. This suggests that circulation statistics is better suited for a universal description of quantum turbulence~\cite{Iyer2019,Polanco2021}. Moreover, based on the scale-invariance of the circulations, we establish a connection with critical models in 2D. Real space analysis allows the identification of correlated vorticity regions with typical size ranging from $\xi$ to $10\xi$. We find that these regions share the same statistical behavior of critical percolation boundaries, enabling a series of predictions to be made on the properties of the inverse energy cascade in 2D quantum fluids. Optical systems provide a new experimental platform that can help unify the microscopic nature of vortex interactions with the large-scale symmetries of turbulent fluids~\cite{caputo2019, liberal2020, maitre2021, eloy2021, ferreira2022, li2022, abobaker2023, bai2023, xiong2023}.
\newpage

\begin{acknowledgments}
The authors thank Guido Boffetta for useful discussions. The authors acknowledge the following projects: Italian Ministry of University (MUR)  PRIN project ``Interacting Photons in Polariton Circuits''– INPhoPOL (grant 2017P9FJBS);
the project ``Hardware implementation of a polariton neural network for neuromorphic computing'' – Joint Bilateral Agreement CNR-RFBR (Russian Foundation for Basic Research) – Triennal Program 2021–2023;
the MAECI project ``Novel photonic platform for neuromorphic computing'', Joint Bilateral Project Italia-Polonia 2022-2023;
PNRR MUR project ``National Quantum Science and Technology Institute'' - NQSTI (PE0000023);
PNRR MUR project ``Integrated Infrastructure Initiative in Photonic and Quantum Sciences'' - I-PHOQS (IR0000016);
Apulia Region, project ``Progetto Tecnopolo per la Medicina di precisione'', Tecnomed 2 (grant number: Deliberazione della Giunta Regionale n. 2117 del 21/11/2018).
The authors are grateful to P. Cazzato for the valuable technical support during the experiments.
%We also acknowledge the project FISR - C.N.R. ``Tecnopolo di nanotecnologia e fotonica per la medicina di precisione'' - CUP B83B17000010001 and ``Progetto Tecnopolo per la Medicina di precisione'', Deliberazione della Giunta Regionale n. 2117 del 21/11/2018.
\end{acknowledgments}

\appendix

\section{Methods}
To build a statistics for the connected regions of vorticity flux like the one shown in Fig.~\ref{fig:3}, we first compute the curl of the incompressible velocity field $\nabla\times\mathbf{u}(\mathbf{x})$ and then we construct the scalar field $\Gamma_R(\mathbf{x})$, for each image and for $R=[1.4\xi,~1.96\xi,~2.5\xi]$. The highest positive (minimum negative) value for this ensemble of fields $\Gamma_R(\mathbf{x})$ is what we use to set a threshold that is 5\%, 10\%, or 20\% of this value. For each field every positive (negative) value that is greater (smaller) than said threshold is set to 1 (-1) while the rest is set to 0, resulting in the ``binarized'' images like the one in Fig.~\ref{fig:3}a.

To analyse the resulting connected regions we labelled them with the python \textit{skimage.measure.label} module, using a 2-connectivity. The area of each region is the number of pixels of the region scaled by the pixel area, while the perimeter is approximated with a line through the centers of border pixels using a 4-connectivity.

\bibliography{biblio}

\end{document}

\preprint{AIP/123-QED}

\title{Supplementary information for: Conformal-invariance of 2D quantum turbulence in an exciton-polariton fluid of light}
%Self-similar properties of 2D quantum turbulence in the inverse energy cascade of exciton-polariton fluids of light}
% Force line breaks with \\
\author{R. Panico}
\affiliation{CNR NANOTEC, Institute of Nanotechnology, Via Monteroni, 73100 Lecce, Italy}

\author{A. S. Lanotte}
\affiliation{CNR NANOTEC, Institute of Nanotechnology, Via Monteroni, 73100 Lecce, Italy}
\affiliation{INFN, Sez. Lecce, 73100 Lecce, Italy}

\author{D. Trypogeorgos}
\affiliation{CNR NANOTEC, Institute of Nanotechnology, Via Monteroni, 73100 Lecce, Italy}

\author{G. Gigli}
\affiliation{CNR NANOTEC, Institute of Nanotechnology, Via Monteroni, 73100 Lecce, Italy}
\affiliation{Dipartimento di Matematica e Fisica E.~De Giorgi, Universit\`a del Salento, Campus Ecotekne, via Monteroni, Lecce 73100, Italy}

\author{M. De Giorgi}
\affiliation{CNR NANOTEC, Institute of Nanotechnology, Via Monteroni, 73100 Lecce, Italy}

\author{D. Sanvitto}
\affiliation{CNR NANOTEC, Institute of Nanotechnology, Via Monteroni, 73100 Lecce, Italy}

\author{D. Ballarini}
\email{dario.ballarini@nanotec.cnr.it}
\affiliation{CNR NANOTEC, Institute of Nanotechnology, Via Monteroni, 73100 Lecce, Italy}

\date{\today}

\maketitle

\section{Compressibility}
All the observables presented in the main discussion are derived from the incompressible part of the density weighted velocity field $\mathbf{u}(\mathbf{x},t)= \sqrt{\rho}(\mathbf{x},t)\mathbf{v}(\mathbf{x},t)$ where ${\rho}(\mathbf{x},t)$ is the polariton fluid density and $\mathbf{v}(\mathbf{x},t)=\frac{\hbar}{m}\nabla\theta(\mathbf{x},t)$ its velocity (see Methods of \citet{panico2023onset}). The superfluid itself however is also made of a compressible component and both contribute to the dynamics. We can quantify the degree of compressibility of our fluid using the standard definition ${\cal P} \equiv {\cal C}^2/{\cal S}^2$, where ${\cal C}$ quantifies the compressible part with respect to the velocity gradient modulus, and $0 \le {\cal P} \le 1$ by definition,
\begin{equation}
    {\cal P} = \frac{\langle\left(\partial_x u_x + \partial_y u_y\right)^2\rangle}
    {\langle\left(\partial_x u_x\right)^2 + \left(\partial_y u_x\right)^2 + \left(\partial_x u_y\right)^2 + \left(\partial_y u_y\right)^2\rangle}\,.
\end{equation}
In our case, it yields ${\cal P}\simeq0.43$. 
%meaning our superfluid is far from what is defined as ``weakly %compressible''. 
%Despite this, our results are in good agreement with theoretical %predictions made for this particular class of %superfluids~\cite{puggioni2020conformal}.

\section{Second order moment of longitudinal velocity increments $\delta_r u$}
\begin{figure}
    \centering
    \includegraphics[width=\columnwidth]{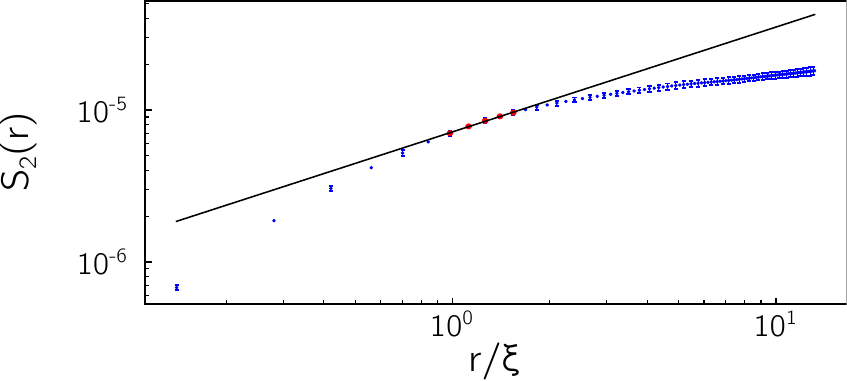}
    \caption{Log-log plot of the second order moment of the incompressible velocity increments $\delta_r u \equiv\frac{\mathbf{r}}{r}\cdot\left[\mathbf{u}(\mathbf{x}+\mathbf{r})-\mathbf{u}(\mathbf{x})\right]$ as a function of the scale. A scaling behaviour $S_2(r) \sim r^{\zeta_2}$ is expected within the turbulent scaling range: a slope is fitted in the scaling region of few healing lengths, which results in the value of the scaling exponent $\zeta_2=0.67\pm0.03$. 
    %c) Values of the exponents $\zeta_p$, obtained as in (a) for various moments $S_p(r)$, in the same spatial interval, show a deviation from self-similarity (dashed curve).
    }
    \label{figSI:1}
\end{figure}

To get further insight in the role of quantised vortices, we look at the scaling behaviour of the p-th order moments of the distributions of the velocity increments reported in the main text as a function of the scale $r$,
\begin{equation}
    S_{p}\left(r\right)=\left<\left|\frac{\mathbf{r}}{r}\cdot\left[\mathbf{u}(\mathbf{x}+\mathbf{r})-\mathbf{u}(\mathbf{x})\right]\right|^{p}\right>\,.
\end{equation}
These exhibit a self-similar behaviour in the classical 2D inverse cascade, $S_{p}(r)\propto S_{3}^{p/3}(r)\sim r^{\frac{p}{3}}$~\cite{boffetta2000inverse}. Here we focus on the second-order moment $S_{2}(r)$ which is shown in Fig.~\ref{figSI:1}, together with the fitted slope corresponding to the range $\xi<r<2\xi$. Despite a small scaling region, it is possible to estimate the scaling exponent as $0.67\pm0.02$, which corresponds to the Kolmogorov-like $-5/3$ scaling law of the energy spectrum in the inverse cascade according to the Wiener–Khinchin theorem. We note that turbulent scaling range is limited on the one hand by the presence on velocity jumps due to the quantised vortices, and on the other hand by the lack of stationarity at large scales.
%As shown in Fig.~\ref{figSI:1}b, the scaling exponents (taken for the same range of $r$) do not follow exactly the expected self-similar behaviour observed in classical 2D turbulence, instead they show a slight anomalous behaviour for $p>3$. We argue that such small anomalous correction might be due to the interplay of different scaling behaviours in a small range of scales. 

\section{Non turbulent case}
\begin{figure}[ht]
    \centering
    \includegraphics[width=\columnwidth]{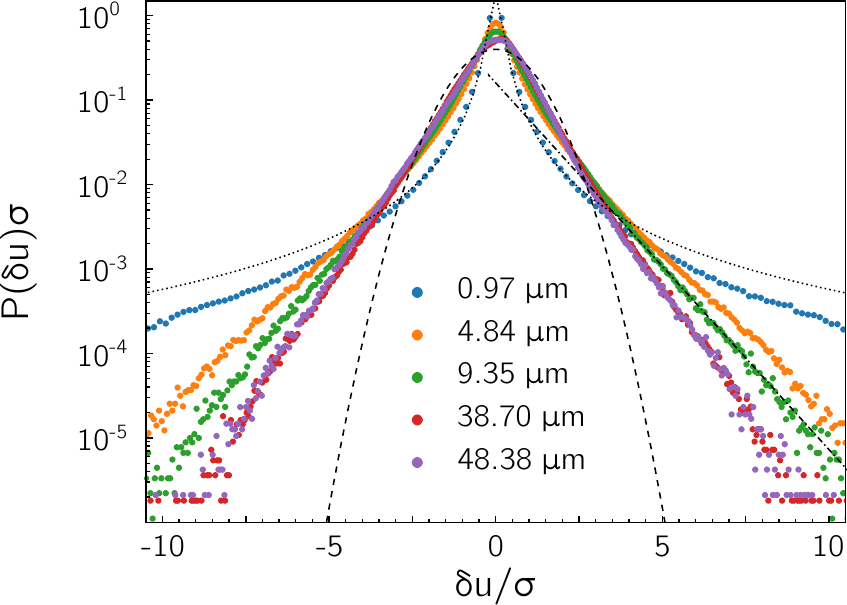}
    \caption{a) Pdfs of the velocity increments $\frac{\mathbf{r}}{r}\cdot\left[\mathbf{u}(\mathbf{x}+\mathbf{r})-\mathbf{u}(\mathbf{x})\right]$ for different $\mathbf{r}$, normalised with their respective standard deviation $\sigma$. As for the turbulent case for sub-healing length scales the pdf follows a Cauchy-Lorentz distribution (dotted line), while increasing $\mathbf{r}$ it slowly approaches a Gaussian (dashed line).}
    \label{figSI:2}
\end{figure}

\begin{figure}[ht!]
    \centering
    \includegraphics[width=\columnwidth]{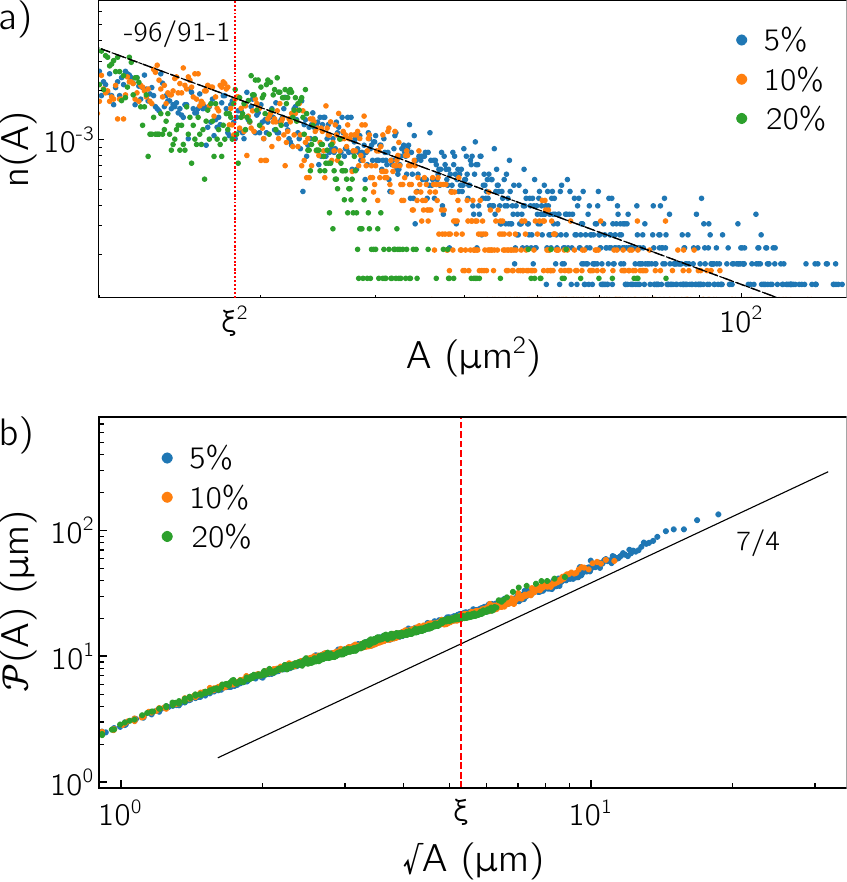}
    \caption{a) Distribution of the extension of the connected area identified by thresholding the circulation. For the non turbulent case the points do not follow a well defined scaling. b) Fractal dimension of the connected regions computed as the perimeter of each region versus their size. The solid black line represent the 7/4 scaling expected for a critical percolation, the dashed line indicates the actual scaling of the non turbulent case.}
    \label{figSI:3}
\end{figure}

We analyse the same system in the main text but for initial conditions that do not lead to a turbulent state~\cite{panico2023onset}. The statistics is constructed on a similar number of points. The normalised PDFs of the velocity increments in Fig.~\ref{figSI:2} are not self-similar, as expected given the quantised structure of the vortices, and exhibit a vortex-dominated behaviour as in the turbulent case illustrated in the main text. When the dimensionality of the system is measured using the same methods discussed in the main book, a noticeable difference emerges. In Fig.~\ref{figSI:3}a, we report the number of connected regions of $\Gamma(\mathbf{x})$ with a given area $n(A)$, demonstrating how the points are much more spread and the scaling - if present - is slower than $\alpha=96/91 +1$. This indicates that the system has a lower tendency to arrange its structures over larger spatial scales. This is again confirmed by the fractal dimension measured as the perimeter $\mathcal{P}(A)$ versus the square root of the area $A$ of the connected regions, as shown in Fig.~\ref{figSI:3}b. In this case, for values above $\xi$, we can observe a scaling slightly lower than the one reported in Fig.~4b of the main text.
It is also worth noting that $\xi$ in this case differs from the one mentioned in the main text, and is of the order of $\xi\simeq5.3$~\si{\micro\meter}. While the connected regions that form are clearly smaller and more scattered than in the turbulent case, the system begins to exhibit similar behaviour, and given enough time (limited by the polariton lifetime), it would most likely reach a very similar distribution of connected regions.

\nocite{*}
% \bibliography{aipsamp}% Produces the bibliography via BibTeX.
\bibliography{biblio_sup}